\documentclass[reprint,amsmath,amssymb,aps,prl,superscriptaddress,longbibliography]{revtex4-2}
\usepackage{graphicx}
\usepackage{xcolor}
\usepackage{bm}
\usepackage{hyperref}
\hypersetup{colorlinks,linkcolor=blue!50!black,citecolor=blue!50!black,urlcolor=blue!50!black}

\begin{document}

\title{Photoelectrons resolve the phase-dependent variance of bright squeezed vacuum}

\author{Haoran Ma}
\affiliation{Department of Physics, Stanford University, Stanford, California 94305, USA}
\affiliation{Stanford PULSE Institute, SLAC National Accelerator Laboratory, Menlo Park, California 94025, USA}
\author{Yihe Wang}
\affiliation{Department of Electrical and Computer Engineering, University of Arizona, Tucson, Arizona 85721, USA}
\affiliation{Department of Physics, University of Arizona, Tucson, Arizona 85721, USA}
\author{Chuan Cheng}
\affiliation{Department of Physics, Stanford University, Stanford, California 94305, USA}
\affiliation{Stanford PULSE Institute, SLAC National Accelerator Laboratory, Menlo Park, California 94025, USA}
\affiliation{State Key Laboratory for Mesoscopic Physics and Frontiers Science Center for Nano-Optoelectronics, School of Physics, Peking University, Beijing 100871, China}
\author{Ian Gabalski}
\affiliation{Stanford PULSE Institute, SLAC National Accelerator Laboratory, Menlo Park, California 94025, USA}
\affiliation{Department of Applied Physics, Stanford University, Stanford, California 94305, USA}
\affiliation{Department of Chemistry, University of California, Davis, Davis, California 95616, USA}
\author{Aaron M. Ghrist}
\affiliation{Stanford PULSE Institute, SLAC National Accelerator Laboratory, Menlo Park, California 94025, USA}
\affiliation{Department of Applied Physics, Stanford University, Stanford, California 94305, USA}
\author{Eleanor Weckwerth}
\affiliation{Department of Physics, Stanford University, Stanford, California 94305, USA}
\affiliation{Stanford PULSE Institute, SLAC National Accelerator Laboratory, Menlo Park, California 94025, USA}
\author{Mohammed Th.~Hassan}
\affiliation{Department of Physics, University of Arizona, Tucson, Arizona 85721, USA}
\affiliation{James C. Wyant College of Optical Sciences, University of Arizona, Tucson, Arizona 85721, USA}
\author{Philip H. Bucksbaum}
\affiliation{Department of Physics, Stanford University, Stanford, California 94305, USA}
\affiliation{Stanford PULSE Institute, SLAC National Accelerator Laboratory, Menlo Park, California 94025, USA}
\affiliation{Department of Applied Physics, Stanford University, Stanford, California 94305, USA}

\date{\today}

\begin{abstract}
The variance of bright squeezed vacuum (BSV) is not only strong but also phase dependent, yet recent strong-field experiments with BSV have recorded only its magnitude. A coherent field of the same color provides the phase reference that resolves this phase-dependent noise. Strong-field ionization of xenon by this light shows that the photoelectron yield follows the BSV quadrature along the coherent field, and that the nonlinearity of ionization amplifies the quadrature noise many times over in the electron yield. The noise is moreover distributed unevenly over the photoelectron momentum distribution, differing by an order of magnitude between ionization channels. Strong-field ionization thus offers a phase-sensitive, channel-resolved detector of quantum light, while the quadrature noise of quantum light offers a probe of the intensity dependence of individual ionization pathways.

\medskip
\end{abstract}

\maketitle

Bright squeezed vacuum (BSV) differs from a coherent optical field not only in the magnitude of its noise but also in its phase dependence: the noise is reduced in one quadrature and amplified in the other \cite{Iskhakov2012,Chekhova2015}. Recent experiments have shown that the strong photon-number noise of BSV modifies strong-field processes, such as high-harmonic generation \cite{Rasputnyi2024}, photoemission from metal needle tips \cite{Heimerl2024,Heimerl2025}, and atomic ionization \cite{Jiang2026,LiuXe2026}. These measurements, along with theory \cite{Gorlach2020,Gorlach2023,Lewenstein2021,Fang2023,RiveraDean2025,RiveraDean2026,Stammer2026,Schimmoller2026,LiuQHHG2026}, establish that highly nonlinear strong-field processes are sensitive to the photon statistics of the driving light, but they do not show how the phase-dependent noise affects the process under study. The BSV field itself has been characterized by single-shot and sub-cycle optical metrology \cite{Kern2026,GholamMirzaei2026}, and electron-based schemes for its noise have been proposed \cite{Stammer2026,Zhou2026}. 

Strong-field ionization should be particularly sensitive to such phase-dependent fluctuations: the quadrature in phase with the field changes its amplitude, the quadrature out of phase shifts the field within the optical cycle, and electron emission and the subsequent dynamics in above-threshold ionization, high-harmonic generation, and nonsequential multiple ionization depend on both.

Here we superpose the BSV with a coherent field of the same color as a phase reference, and show that the photoelectrons of strong-field ionization read out the BSV quadrature along it. The relative phase of the two fields selects which BSV quadrature adds to the coherent field: with the squeezing axis parallel or perpendicular to the coherent field, the noise enters its amplitude or its phase, and the superposition is phase-squeezed or amplitude-squeezed displaced BSV, the squeezed coherent light whose strong-field ionization and harmonic emission have so far been studied theoretically \cite{Habibovic2026,Sennary2026}. We tag the BSV field of every shot optically and sort the shots by this phase, so that the same light is measured as phase-squeezed and as amplitude-squeezed displaced BSV. The photoelectron yield and its fluctuations are compared between the two, and the fluctuations are then resolved over the photoelectron spectrum, where they single out the ionization channels by their intensity dependence.

\begin{figure*}
\includegraphics[width=\textwidth]{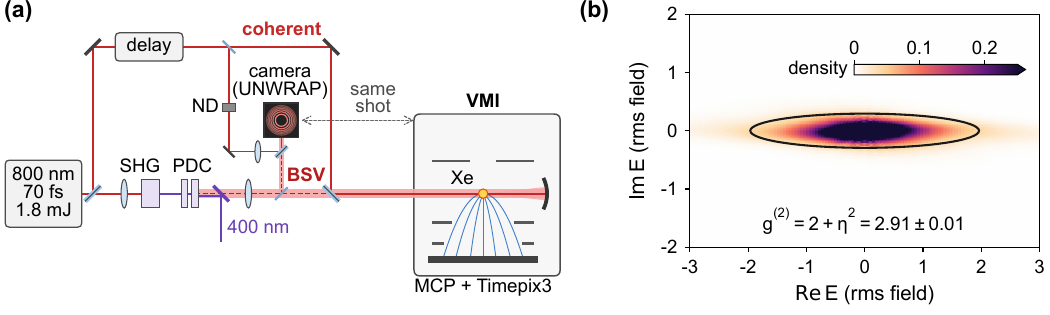}
\caption{\label{fig1}Experimental arrangement and optical characterization of the BSV field. (a) In the BSV arm, second-harmonic generation (SHG) in a BBO crystal pumps parametric down-conversion (PDC) in two BBO crystals; the 400-nm pump (violet) is then removed by a dichroic mirror. The BSV and the coherent field are combined and focused in the velocity-map imaging (VMI) spectrometer \cite{Gabalski2026,Weckwerth2026,Cheng2026water}, which is filled with xenon at low pressure; pick-offs of both arms form Newton rings on a camera, which give the BSV amplitude and phase on every shot (UNWRAP) \cite{Ma2026unwrap}. (b) Husimi function of the BSV field sampled by the camera, density relative to its maximum (linear, saturated above 0.25); real and imaginary parts of the field in units of its rms magnitude. The black ellipse has semi-axes $2a$ and $2b$, twice the rms widths $a = 0.98$ and $b = 0.147$ along the noisy and quiet axes. For a Gaussian field the fourth-moment theorem \cite{Isserlis1918} gives the intensity correlation from the shape alone \cite{Leuchs2015}, $g^{(2)} = 2 + \eta^2$ with $\eta = (a^2-b^2)/(a^2+b^2)$, 2 for a circular (thermal) and 3 for a line-like (squeezed) distribution; here $g^{(2)} = 2.91 \pm 0.01$, a lower bound since $b$ still contains the camera noise.}
\end{figure*}

Figure~\ref{fig1}(a) shows the setup. The output of a Ti:sapphire amplifier is split into two beams. One beam generates the BSV by second-harmonic generation (SHG) followed by degenerate parametric down-conversion (PDC) \cite{Spasibko2017,Manceau2019}, and the BSV is then combined with the other beam, which serves as the coherent field. Pick-offs from both arms form Newton rings on a camera, from which the BSV amplitude and phase relative to the coherent field are retrieved on every shot by Newton-ring interferometry (UNWRAP) and paired with the electrons of the same shot. The BSV generation and the ring readout are described in Ref.~\cite{Ma2026unwrap}. The combined beam is focused into a velocity-map imaging (VMI) spectrometer filled with xenon; an MCP--Timepix3 detector records the electron momentum distribution. The spectrometer is described in Refs.~\cite{Gabalski2026,Weckwerth2026,Cheng2026water}. The relative phase of the BSV and the coherent field is not controlled but drifts over the run, which scans the squeezing axis relative to the coherent field; the phase offset between camera and target is tracked with an out-of-sample fit to the electron yield. On each shot the camera gives the BSV amplitude $A$, in units of its rms value, and its phase $\psi_t$ relative to the coherent field at the target. The two quadratures of the BSV field, parallel and perpendicular to the coherent field, are its real and imaginary parts, $\mathrm{Re}\,E=A\cos\psi_t$ and $\mathrm{Im}\,E=A\sin\psi_t$; $\mathrm{Re}\,E>0$ means that the BSV field adds to the coherent field. On each shot $N$ photoelectrons are detected. The yield $Y(A,\psi_t)$ is the mean of $N$ over all shots with the same measured field.

Figure~\ref{fig1}(b) shows the BSV field distribution measured by the camera. The camera records the amplitude and the phase of each shot, and their joint distribution is the Husimi function of the BSV \cite{Rasputnyi2026}. The phase jitter between the two arms, about $10^\circ$ rms, has been removed from it by deconvolution. The distribution is elongated along one axis, the squeezing axis, with equal populations of the two field signs, and the rms widths of the noisy and quiet quadratures differ by a factor of 6.7 (4.4 before the deconvolution); the quiet width still contains the camera noise, so the ratio is a lower bound. We sort shots by the measured phase $\psi_t$, folded into an angle $0\leq\phi\leq90^\circ$ between the BSV field and the coherent field, since $\psi_t$ and $\psi_t+\pi$ belong to the same squeezed state. The relative phase of the two arms drifts slowly, so over the run the squeezing axis of the state in Fig.~\ref{fig1}(b) sweeps through all orientations relative to the coherent field. The measured angle of a shot tells where the axis was at that moment, and sorting by it sorts the shots by the orientation of the axis. The field of a single shot is a distribution, which we replace by the one value the camera records; the camera resolution is broader than the quiet width of the state \cite{Ma2026unwrap}, so nothing is lost. Shots with $\phi<20^\circ$ are therefore phase-squeezed light, with the squeezing axis parallel to the coherent field, and shots with $\phi>70^\circ$ amplitude-squeezed light, with the axis perpendicular to it. As a reference we use all BSV shots together, without sorting by $\phi$; in this random-phase sample the squeezing axis takes all orientations. 

Within each selection the shots are binned by their measured field $(A,\psi_t)$, the field-correlated noise is estimated as
\begin{equation}
\frac{\mathrm{Var}[Y]}{\langle N\rangle^2}\equiv g^{(2)}-1=\frac{\mathrm{Var}_k(\lambda_k)-\langle(1-w_k)\,\sigma_k^2/n_k\rangle_k}{\langle N\rangle^2}.
\label{eq:g2}
\end{equation}
The left-hand side is the quantity of interest: the variance of the yield $Y(A,\psi_t)$ over the field distribution, normalized by the squared mean count $\langle N\rangle$ of the selection. The right-hand side is our observable. Bin $k$ contains $n_k$ shots, with sample mean $\lambda_k$ and sample variance $\sigma_k^2$ of the electron count; $\mathrm{Var}_k$ and $\langle\cdot\rangle_k$ are the variance and the mean over bins, each weighted by $w_k=n_k/\sum_k n_k$. The variance of $\lambda_k$ over bins alone would overestimate $\mathrm{Var}[Y]$: each $\lambda_k$ is estimated from a finite number of shots and scatters by $\sigma_k^2/n_k$ around the true yield. This sampling term is large because few electrons are detected per shot, and it carries the coherent-pulse and gas-density fluctuations that are uncorrelated with the field. Subtracting it makes the numerator of Eq.~(\ref{eq:g2}) an unbiased estimate of the variance of the bin-averaged yield. The uncertainties for $g^{(2)}-1$ are standard deviations over bootstrap resamples of the 60-s data blocks.

\begin{figure*}
\includegraphics[width=\textwidth]{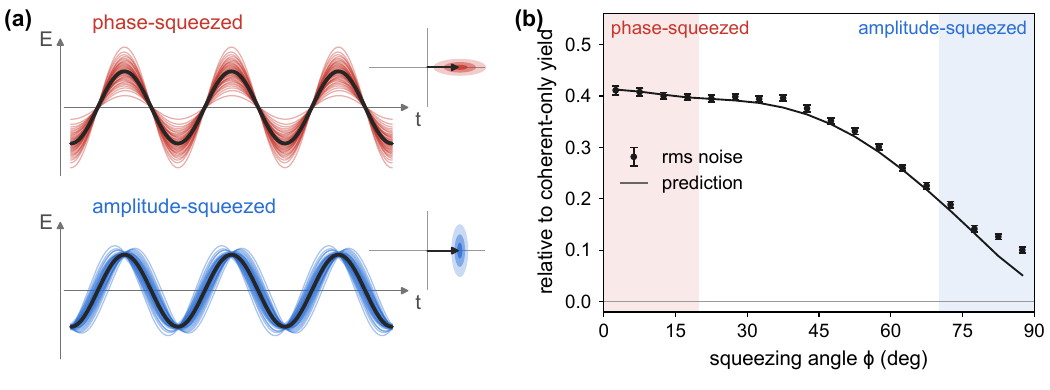}
\caption{\label{fig2}Phase-squeezed and amplitude-squeezed displaced BSV, and the photoelectron yield as a function of the squeezing angle. (a) Sketch of the electric field: the coherent field (black) and the total field of individual shots. For phase-squeezed BSV (red, top) the shots differ in amplitude and spread at the crests; for amplitude-squeezed BSV (blue, bottom) they are shifted in phase and spread at the zero crossings. Insets: the same two cases in phase space. (b) Rms noise of the photoelectron yield relative to the coherent-only yield (BSV amplitude $A<0.3$), against the squeezing angle $\phi$; error bars are bootstrap standard deviations. Solid line: prediction of the noise for displaced BSV, whose intensity noise follows the quadrature along the coherent field. Shaded: the phase-squeezed ($\phi<20^\circ$) and amplitude-squeezed ($\phi>70^\circ$) selections.}
\end{figure*}

Figure~\ref{fig2}(a) sketches two cases. With the squeezing axis parallel to the coherent field, the BSV adds to or subtracts from it, so the shots differ in amplitude while their zero crossings stay fixed: the light is phase-squeezed. With the squeezing axis perpendicular to the coherent field, the BSV shifts the zero crossings and leaves the amplitude almost unchanged: the light is amplitude-squeezed.

Figure~\ref{fig2}(b) shows the field-correlated rms noise of the yield against the squeezing angle $\phi$, relative to the coherent-only shots ($A<0.3$). The excess noise falls from about 40\% for phase-squeezed light to about 10\% for amplitude-squeezed light. This is the behaviour expected of displaced BSV, shown by the line in Fig.~\ref{fig2}(b): the contribution of the BSV to the intensity variance follows $\langle(\mathrm{Re}\,E)^2\rangle=\langle A^2\cos^2\phi\rangle$, which falls from its maximum to its minimum over $90^\circ$. The camera measures this quadrature directly. Relative to all shots, its second moment $\langle(\mathrm{Re}\,E)^2\rangle$ is larger by a factor of $1.54\pm0.04$ in the phase-squeezed selection and smaller by a factor of $0.121\pm0.005$ in the amplitude-squeezed selection, while the electron variance $g^{(2)}-1$ changes by factors of $1.44\pm0.03$ and $0.19\pm0.01$, respectively. The optical and the electron factors agree qualitatively, so the quantity the electrons measure is dominated by $\mathrm{Re}\,E$, the BSV quadrature along the coherent field.

\begin{figure*}
\includegraphics[width=\textwidth]{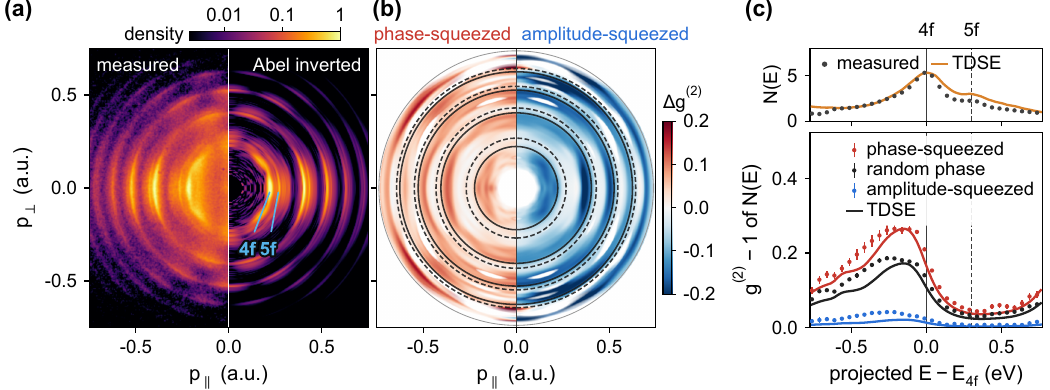}
\caption{\label{fig3}Momentum- and energy-resolved photoelectron fluctuations correlated with the BSV field. (a) Xenon photoelectron momentum distribution, shown on a logarithmic color scale: distortion-corrected VMI image (left) and Abel inversion (right), each normalized to its maximum. The laser polarization is horizontal. The 4f and 5f Freeman resonances are marked in the first ATI order. (b) $\Delta g^{(2)} = g^{(2)} - g^{(2)}_{\rm random}$, the change of $g^{(2)}-1$ [Eq.~(\ref{eq:g2}), per pixel] in the phase-squeezed (left, $\phi < 20^\circ$) and amplitude-squeezed (right, $\phi > 70^\circ$) selections relative to all BSV shots. Solid and dashed circles: 4f and 5f peaks. (c) Angle-integrated spectrum $N(E)$ of the measured image (top, $10^{-2}$ electrons per shot per eV) and $g^{(2)}-1$ of $N(E)$ for the three selections (bottom), ATI orders 1--5 folded onto one order, energy measured from each order's projected 4f peak. Points: measurement, with $\pm1\sigma$ bootstrap error bars. Lines: TDSE calculation for the measured pulse, focal-volume averaged and projected, the spectrum normalized to the measured peak and the noise from Eq.~(\ref{eq:tdse}) with $\varepsilon=2.9\%$.}
\end{figure*}

Figure~\ref{fig3} resolves the noise over the photoelectron distribution. At a coherent peak intensity of $2.0\times10^{13}\,\mathrm{W\,cm^{-2}}$ ($U_p=1.2$\,eV, Keldysh parameter $\gamma=2.2$ \cite{Keldysh1965}) the mean spectrum consists of ATI peaks spaced by one photon energy \cite{Agostini1979}, of which the first five orders are shown in Fig.~\ref{fig3}(a). Each order has its main peak at the eight-photon Freeman resonance with the 4f state, a 5f satellite 0.31\,eV higher, and non-resonant electrons on its low-energy side \cite{Freeman1987,Schyja1998}. The maps in Fig.~\ref{fig3}(b) show the field-correlated noise of every pixel relative to the random-phase reference. For phase-squeezed light the noise exceeds the reference on every ATI ring; for amplitude-squeezed light it falls below it on every ring. The noise is concentrated on the rings themselves, where the electrons are, and the difference between the two selections grows with the order, as expected when a higher order depends more steeply on the intensity \cite{LiuXe2026}. Within an order, however, the noise is not uniform. Figure~\ref{fig3}(c) looks inside a peak: ATI orders 1--5 are folded onto a single order \cite{Cheng2020} and the noise is plotted against the energy within the order. It is largest for the non-resonant electrons, falls across the 4f peak, and almost vanishes at the 5f satellite.

The BSV field is known from the camera: the intensities of the two pick-offs give an rms BSV field of $\varepsilon\approx2.6\%$ of the coherent field amplitude, uncertain by about a quarter through the attenuation of the coherent pick-off. This small field has a large effect on the electrons: it changes the total yield by a third, rms. After Abel inversion of the image, the field-correlated noise $g^{(2)}-1$ of the random-phase selection is $0.51\pm0.03$ for the non-resonant electrons, $0.17\pm0.01$ at the 4f peak and $0.015\pm0.011$ at the 5f satellite; the non-resonant and the 5f electrons differ by more than an order of magnitude.

A time-dependent Schr\"odinger equation (TDSE) calculation reproduces these features. We use the QPROP code \cite{Bauer2006,Tulsky2020} for a single active electron of xenon in a local potential of the Tong--Lin form \cite{TongLin2005}, refitted to the xenon Rydberg levels \cite{NIST_ASD}, calculate the photoelectron spectrum for the measured pulse, both fine-structure channels of the ion and $|m_l|=0,1$, over a range of peak intensities, and average the spectra over a Gaussian focal volume. The calculated spectrum reproduces the ATI orders, the 4f peak with its 5f shoulder and the non-resonant electrons [Fig.~\ref{fig3}(c), top], and confirms the assignment of the peaks. The response to the BSV then follows with a single free parameter, the BSV field $\varepsilon$. The BSV field adds to the coherent field, so the intensity of a shot is $I_0\,|1+\varepsilon E|^2=I_0\,(1+2\varepsilon\,\mathrm{Re}\,E+\varepsilon^2A^2)$. Near $I_0$ the calculated yield at each energy scales as a power of the intensity, $Y(E)\propto I^{S(E)}$, so to first order the yield of a shot changes by the fraction $S(E)\,(2\varepsilon\,\mathrm{Re}\,E+\varepsilon^2A^2)$, and its variance over the shots of a selection is
\begin{equation}
g^{(2)}-1=S(E)^2\,\mathrm{Var}\!\left[2\varepsilon\,\mathrm{Re}\,E+\varepsilon^2A^2\right],
\label{eq:tdse}
\end{equation}
with the variance taken over the camera field of each selection. The lines in Fig.~\ref{fig3}(c) use $\varepsilon=2.9\%$, which qualitatively agrees with the camera value. The calculation reproduces the maximum of the noise for the non-resonant electrons, its minimum at the 5f satellite and the ordering of the three selections.

The calculated power $S(E)$ is largest for the non-resonant electrons on the low-energy side of each order, intermediate at the 4f peak and smallest at the 5f satellite, and the three channels respond to the intensity in three different ways. The non-resonant electrons originate near the peak intensity, where the ionization rate rises most steeply with intensity, and their energy shifts with $U_p$, so both their weight and their energy change with the intensity. A Freeman channel, by contrast, is populated only where the pulse reaches the intensity at which the Stark-shifted Rydberg state comes into resonance, so its energy is fixed and its yield changes mainly through the focal volume that reaches that intensity \cite{Freeman1987,Marchenko2010}; the two resonances then differ by focal geometry \cite{Weckwerth2025}. The ponderomotive shift brings the 4f state into eight-photon resonance at $1.8\times10^{13}$\,W\,cm$^{-2}$ \cite{Schyja1998}, within 10\% of the peak intensity, so the 4f electrons come from a small volume around the focal center, which a small increase of the peak intensity enlarges strongly. The 5f state, bound 0.31\,eV less tightly, comes into resonance at about $1.3\times10^{13}$\,W\,cm$^{-2}$, in a much larger volume, which the same increase enlarges only slightly; in the TDSE calculation the 4f and 5f peaks indeed appear at these two intensities. In ionization by BSV alone the yield fluctuations grow with the electron energy, because the most energetic electrons come from the rare, most intense shots \cite{LiuXe2026}. Here the coherent field sets the intensity and the BSV changes it by only a few percent, so the fluctuations measure how steeply each channel depends on the intensity at the working point. From one ATI order to the next this dependence grows with the number of absorbed photons; within an order it follows the channel, and the contrast between the non-resonant and the resonant electrons is far larger than the change between orders.

In summary, the photoelectron yield of xenon follows the BSV quadrature along the coherent field, so that the relative phase of the two fields decides how much of the BSV noise reaches the electrons, and this noise, resolved over the spectrum, measures the intensity dependence of each ionization channel. What the electrons have registered is the amplitude noise of the displaced BSV: anti-squeezed when the squeezing axis is parallel to the coherent field, squeezed when it is perpendicular. The phase noise of the light, squeezed and anti-squeezed in the opposite order, has not shown itself here. Strong-field ionization is, however, inherently sensitive to the phase of the field: the electron is released near a crest of the field, and its subsequent motion is set by the phase at which it was released \cite{Corkum1993,Paulus2001,Baltuska2003,Krausz2009}. The phase noise of the BSV should therefore become visible in the electrons as well, and the atom would then read out both quadratures of the quantum light.

\begin{acknowledgments}
This work was funded by the UK Research and Innovation--National Science Foundation joint program, ``Exploiting Quantum Information Science in Chemistry''. H.M., Y.W., M.H., and P.B. were supported from this program. C.C., A.G., and E.W. were supported by the National Science Foundation, PHY-2309238.  I.G. was supported by the Department of Energy BES CSGB AMOS program. We acknowledge critical discussions with the APIQuL collaboration, especially Carla F.d.M. Faria, Philipp Stammer, and Javier Rivera-Dean.
\end{acknowledgments}

\bibliography{refs}

\end{document}